# Photonic hook plasmons: a new curved surface wave

**Igor V. Minin,[1,*] Oleg V. Minin,[2] Igor A.Glinskiy,[3,4] and Dmitry S.Ponomarev [3,4]**

[1]*National Research Tomsk Polytechnic University, Lenin Ave., 30, Tomsk, 634050, Russia*
[2]*National Research Tomsk State University, Lenin Ave., 36, Tomsk, 634050, Russia*
[3]*Institute of Ultra High Frequency Semiconductor Electronics RAS, Moscow 117105, Russia*
[4]*Prokhorov General Physics Institute of RAS, Moscow 119991, Russia*
*\*Corresponding author: ivminin@tpu.ru*

**It is well-known that surface plasmon wave propagates along a straight line, but this common sense was broken by the artificial curved light – plasmon Airy beam. In this paper, we introduce a new class of curved surface plasmon wave - the photonic hook plasmon. It propagates along wavelength scaled curved trajectory with radius less than surface plasmon polariton wavelength, and can exist despite the strong energy dissipation at the metal surface.**

Plasmonic structures have been intensively studied due to the ability to squeeze light into objects with sub-wavelength size thanks to the nature of surface plasmon polaritons (SPP) [1]. Plasmons are essentially two-dimensional waves whose field components decay exponentially with distance from the surface. The very fact that these waves tightly cling to the surface makes them ideal for biosensing applications, micromanipulations [2] and molecule diagnostics [3].

In recent years, SPPs beam has been discovered in novel forms with nondiffracting properties rather than the common beam, realized in the in-plane of the interface as classical light in a free space. The investigations of SPPs by using the zeroth-order Bessel beams was reported in [4], the existence of nondiffracting Bessel SPP was studied in [5]. In a planar system, however, there are only a few choices for the construction of curved beams. Airy beams as a solution of the paraxial Helmholtz equation [6] have been suggested theoretically in plasmonics in [7] and later experimentally verified in [8-12]. Realization of plasmonic Mathieu and Weber beams were also reported [13]. However, in low-dimensional systems, such as plasmonic [14] or graphene [15], in which at least in one of the three dimensions electronic state wave function is confined, today the families of plasmon Airy beams are the only beams that have a curved trajectory. But for Airy-type SPPs high beam acceleration is required to achieve significant curvature over the propagation length.

On the other hand, a new term "photonic nanojet" (PJ) was coined [16,17] for the sub-wavelength-scale near field focusing at the shadow side of a mesoscale dielectric particle. Later the theoretical possibilities of PJ formation at surface plasmon were considered for dielectric flat disc [18] and flat cuboid [19] particles. It was shown that the optimal refractive index contrast lies between 1.3 and 1.75. Moreover, the SPP PJ is moved away from the cuboid as it is enlarged. It was also demonstrated that, even when the SPPs are not strongly coupled to the surface of the metal film, it is possible to produce PJs using dielectric cuboids with a subwavelength focal spot and detect subwavelength ($0.1\lambda$) metal particles [19]. Lately it was shown that the SPP PJ performance can be engineered by simply changing the height of the dielectric at the operation wavelength for the flat disk [18], cuboid [19] and also with the plasmonic lenses [20-21].

Recently, a photonic hook (PH) [22] as a new type of subwavelength curved beam which is based on PJ effect has been observed in a free space [23]. Following the analogy with the process of a free-space in order to realize PH as surface plasmon waves, several fundamental challenges should be addressed due to the plasmonic nature of the waves. First, due to SPPs are *p*-polarized inhomogeneous plane waves [24] the real and imaginary parts of k-wave vector, are not parallel to each other, i.e. require a compensation momentum to match the two wave-vectors of the beams. Second, as it was mentioned above, since the SPPs have a very short propagation distance due to the large Ohmic loss, the resulting beams should be formed directly in the near-field, before they decay [24,25]. Moreover, SPPs are excited over a finite propagation distance and therefore their phase cannot be simply defined at a specific one-dimensional plane [19]. Although it is possible to increase the propagation length of SPP waves using, for example, cascaded dielectric particles [26] or metamaterials [26].

To the best of our knowledge, there is no research published about the formation of curved surface plasmon PJs. In this Letter, we report on a new way to produce the curved SPP beam, named "photonic hook plasmons" (PHP) to emphasize its curvature characteristic. The key difference between PHP and surface plasmon Airy beam is that the first is created using the in-plane focusing of SPP wave through an asymmetric dielectric particle which is the combination of a wedge prism and a flat cuboid. This process is fundamentally simpler than the generation of the SPP Airy-family beams [7-13]. The second key difference is a subwavelength curvature of PHP in comparison to plasmonic Airy beams. The formation of the PH in a free-space [23] and in a symmetrically flat cuboid particle in SPP [19] has been previously investigated. Relying on the established knowledge, the influence of the size and prism angle of an asymmetric particle on the distribution of field enhancement, width (full-width at half-maximum [FWHM]), and curvature of PHP is effectively quantified in this Letter.

A three-dimensional full wave simulation using FEM method in COMSOL MULTIPHYSICS is performed to demonstrate PHP generated by the proposed structure. In the simulation, SPP propagates along the negative x direction and is incident onto the side surface of the cube (with prism apex as shown in Fig.1a). Excited by a light source with wavelength $\lambda_0 = 800$ nm, the illuminating SPP has a maximum electric field $E = 1$ V/m at the dielectric-metal interface. Perfect matched layers with scattering boundaries are used to fully absorb the outward waves. A non-uniform mesh with maximum cell size of $1/5\lambda_0$ at the dielectric-metal interface is applied to guarantee the simulation accuracy as in [18-19]. All the field distributions shown below are normalized by their maximum values.

Refractive index of dielectric cube and prism (with angle θ) is 1.35. We used a Drude-Lorentz dispersion model in simulation with relative permittivity of metal (gold) equal to $\varepsilon_m = -23.4 + i1.55$ at $\lambda_0 = 800$ nm. The effective refractive index of SPP [18,19,21] is determined by both dielectric and metal that form interface: $k_{spp} = k_0(\varepsilon_m \cdot \varepsilon_d)/(\varepsilon_m + \varepsilon_d)1/2 = n_{eff}k_0$, where $k_0 = 2\pi/\lambda_0$ is the wave number in vacuum [21], $\varepsilon_m$ and $\varepsilon_d$ are the relative permittivities of metal and dielectric, respectively. It means that surface plasmon wavelength is $\lambda_{spp} = 0.979\lambda_0 = 783$ nm. The height of the dielectric and gold was $h = 600$ nm and 100 nm, respectively (Fig.1a).

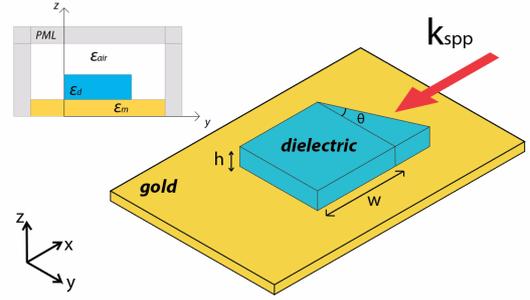

Fig. 1a. Dielectric cube with prism for generation the SPP photonic hook

In the Fig.1 (b-g) SPP field intensity ($E^2$) distributions for dielectric particles with different dimensions and without/with optimal prism are shown.

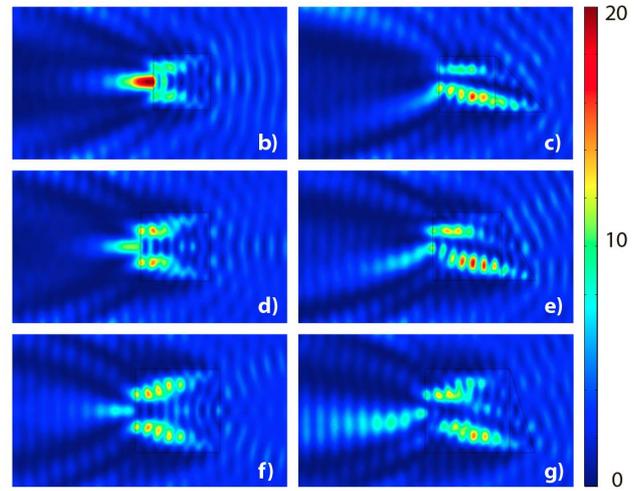

Fig. 1(b-g). SPP field intensity ($E^2$) distributions for different dielectric particles: $2.0\lambda_{spp} \times 2.0\lambda_{spp}$, θ=0 (b), $2.0\lambda_{spp} \times 2.0\lambda_{spp}$, θ=44.4⁰ (c), $2.5\lambda_{spp} \times 2.5\lambda_{spp}$, θ=0 (d), $2.5\lambda_{spp} \times 2.5\lambda_{spp}$, θ=23.6⁰ (e), $3.0\lambda_{spp} \times 3.0\lambda_{spp}$, θ=0 (f) and $3.0\lambda_{spp} \times 3.0\lambda_{spp}$, θ=15.5⁰ (g).

One can see in (Fig.1b,d,f) that SPP PJ and PHP are confined at the air-gold interface and an increase in the size of the cube leads to an increase in the length of the plasmonic PJ, but its intensity decreases. Similarly, as the particle size increases for optimal prism sizes, the length of the PHP increases, but its curvature decreases (Fig.1c,e,g).

Figure 2a describes the $E^2$ field enhancement distribution near the shadow surface of particle and FWHM of the plane wave passing through the symmetric and asymmetric particles. An asymmetric particle assembled by a cuboid of the same size and a wedge prism with θ = 23.6° is illuminated by a plane wave, as shown in Fig. 1(a). Here, the prism not only lets the plasmonic PJ shift away from the optical axis, but also bends it in the central propagation, forming a 'hook' shape. It is shown that FWHM of PHP (solid red curve) is shorter than that for plasmonic PJ jet induced by the symmetric cuboid particle (dashed black curve), and is able to break the simplified diffraction limit (0.5λ criterion).

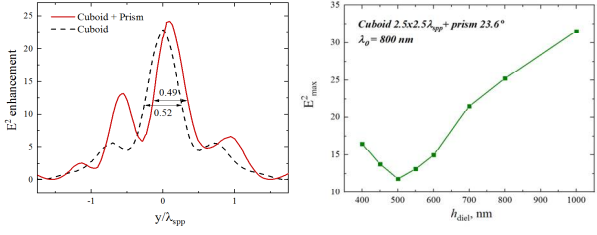

Fig. 2. (a) $E^2$ enhancement profile along $y$ axis for the symmetric cuboid 2.5×2.5$\lambda_{spp}$ without and with prism θ = 23.6°; (b) $E^2_{max}$ enhancement with variation of dielectric height for cuboid 2.5×2.5$\lambda_{spp}$ and θ = 23.6°.

The strong enhancement attributes to two parts: field enhancement at the dielectric-metal interface caused by SPP and field enhancement at the peak amplitude point from SPP focusing due to the SPP PJ effect [18,19] - the electric field of the highly confined PJ at metal-dielectric interface is enhanced by about 20 times for flat cube dimensions of 2$\lambda_{spp}$×2$\lambda_{spp}$ and a height of h=600 nm.

Figure 2b shows $E^2$ enhancement with variation of dielectric height ($h_{diel}$) for cuboid 2.5×2.5$\lambda_{spp}$. As expected, $n_{eff}$ augments as the height $h_{diel}$ increases and it reaches a saturation value for a given thickness (around 1000 nm here), which is related to the exponential decay of SPP electric field within the dielectric medium [21]. The transmitted phase dependence on a cube height becomes close to linear with increasing $h_{diel}$, when the height of the cube exceeds 3 depths of penetration of SPP into the medium [21] and mechanism of PHP formation becomes close to a free-space [23]. For shorter exciting wavelength, the propagation length of PHP dramatically decreases due to the increasing loss of gold film. Thus by varying $h_{diel}$, the formation of PHP can be flexibly controlled similar to SPP PJ [19].

In Fig.3 the profile of PHP and FWHM along curved PHP are shown for different dimensions of dielectric particles and dielectric height of $h_{diel}$ = 600 nm. One can see that the maximal curvature is observed for minimal dimensions of cube (2$\lambda_{spp}$×2$\lambda_{spp}$) and increasing the particle size leads to a decrease in the curvature of the PHP (Fig.3a). The change in FWHM along PHP with increasing particle size has a similar character, while the minimal values of FWHM are 0.465$\lambda_{spp}$, 0.462$\lambda_{spp}$ and 0.408$\lambda_{spp}$ for 2$\lambda_{spp}$×2$\lambda_{spp}$, 2.5$\lambda_{spp}$×2.5$\lambda_{spp}$ and 3$\lambda_{spp}$×3$\lambda_{spp}$ cube dimensions near the shadow surface of particle (Fig.3b), respectively.

Figure 4 describes the effect of critical prism angle to PHP formation for cube with 2.5$\lambda_{spp}$×2.5$\lambda_{spp}$ dimensions. One can see that the angle change influences the propagation trajectory curvature. At a critical inclination θ (more than 44⁰ for this dimensions) the PHP propagates along an oblique straight line (Fig.4d). The diversity of the thickness along $x$ axis with an asymmetric particle produces an unequal phase of the transmitted plane wave which is perpendicular to $y$ axis, resulting in the irregularly concave deformation of the wave front inside particle, leading to the formation of a plasmonic PH.

The offered system has one more interesting property – PHP launched at different wavelengths will propagate along trajectories of different curvatures (Fig.5). In Fig.5a-b the configurations of PHP for cube dimensions of 2.5$\lambda_{spp}$×2.5$\lambda_{spp}$ prism angle 23.6° and illuminated wavelength of 800 nm are shown for different dielectric height. One can see that an increase of the dielectric height leads to concentration of the local field intensity near the shadow surface of particle, hence increasing the intensity of PHP according to Fig.2b.

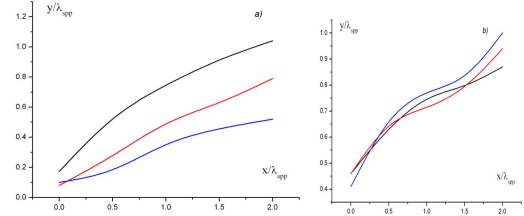

Fig. 3. Curvature of PHP (a) and FWHM along PHP (b) for different particle dimensions: 2.0$\lambda_{spp}$×2.0$\lambda_{spp}$, θ=44.4⁰ (black), 2.5$\lambda_{spp}$×2.5$\lambda_{spp}$, θ=23.6⁰ (red) and 3.0$\lambda_{spp}$×3.0$\lambda_{spp}$, θ=15.5⁰ (blue)

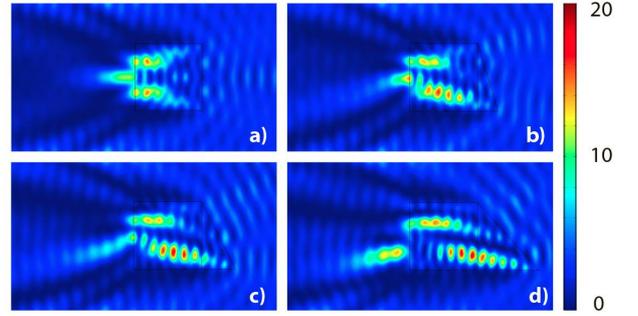

Fig.4. 2D SPP field intensity ($E^2$) map for dielectric particles with 2.5$\lambda_{spp}$×2.5$\lambda_{spp}$ and different prism angles: θ=0°(a); θ=15.5°(b); θ=23.6°(c); θ=44.4°(d).

The change in wavelength causes a change in PHP shape. Comparing the profile of localized SPPs it can be seen (Fig.5e) that for the wavelengths of 700 nm (Fig.5d) and 900 nm (Fig.5c) the profile of the PH is "bent" in a opposite direction.

In conclusion, a novel class of surface plasmons capable of propagating along curved trajectory and called "photonic hook plasmon" is achieved. PHP has a unique property among all other curved surface waves such as subwavelength curvature. In contrast to previous PH in a free-space, our approach is entirely fulfilled in a planar dimension that offers a thoroughly compact manipulation of the plasmonic near-field. It is shown that narrower focus can be offered by this PHP compared to those of SPP PJ produced by normal cuboid lens.

The planar cuboid based structure offers many advantages such as fabrication ease and high integration for on-chip biological or chemical sensing. Such a planar system with a subwavelength confinement is attractive for "flatland photonics" [27] and may find interesting applications in energy routing over plasmonic boards, plasmonic circuitry, in-plane SPP switch and surface tweezers.

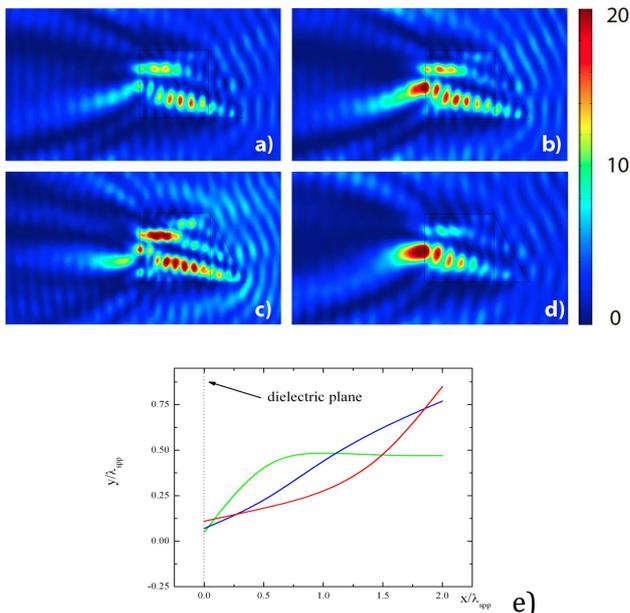

Fig.5. 2D SPP field intensity ($E^2$) map for dielectric particles with $2.5\lambda_{spp}\times 2.5\lambda_{spp}$, θ = 23.6° with dielectric height of 600 nm and illuminated wavelength of 800 nm (a); dielectric height of 1000 nm and illuminated wavelength of 700 (b), 800 (c) and 900 nm (d), respectively; (e) profile of PHP for dielectric height of 1000 nm and illuminated wavelength of 700 (green), 800 (blue) and 900 (red) nm.

Apart from being interesting in their own right, PHPs may also hold promise for new, exciting applications in the general area of plasmonics. In addition, the demonstration of such curved plasmon waves suggests that similar entities can be used in other low-dimensional systems such as graphene and magnetic surfaces. This field is only now beginning.

**Acknowledges.**
D.P. and I.G. were supported by the Russian Scientific Foundation (Project No.18-79-10195). I.M. was partially funded from TPU Competitiveness Enhancement Program grant. I.M and O.M. initiated the ideas of this work and wrote the paper. I.G. performed the numerical simulations. All authors contributed to discussions and editing this paper.